\documentclass[10pt]{article}
\usepackage{amsfonts, amsmath, color, fleqn, graphicx, natbib, setspace,epstopdf, palatino}
\bibpunct{(}{)}{,}{a}{}{;}
\usepackage[hang]{caption}
\usepackage[margin=1in, nohead]{geometry}
\newtheorem{theorem}{Theorem}[section]

\newtheorem{assumption}[theorem]{Assumption}

\newtheorem{corollary}[theorem]{Corollary}

\newtheorem{definition}[theorem]{Definition}

\newtheorem{proposition}[theorem]{Proposition}

\title{Improving Server Utilization in a Distributed Computing Set-up with Independent Clients}
\author{Anindya S. Chakrabarti\thanks {Economics Area, Indian Institute of Management, Vastrapur, Ahmedabad, Gujarat-380015, India, email: anindyac@iima.ac.in.} \ and Diptesh Ghosh\thanks{Production \& Quantitative Methods Area, Indian Institute of Management, Vastrapur, Ahmedabad, Gujarat-380015, India, email: diptesh@iima.ac.in}}

\begin{document}
\maketitle
\begin{abstract}
We consider a set-up in which there are multiple servers and multiple clients in a large distributed computing system. Clients request servers to process jobs. Servers can only process one job in unit time. There is no coordinating agent to route client requests to servers, and clients choose servers independently and simultaneously, and only have access to the outcomes of their own past requests. If more than one clients choose the same server, then only one randomly chosen client's request will be fulfilled. If some servers do not receive any request, they remain idle. In this paper, we show that a large category of strategies are not effective in terms of server utilization. We devise strategies for clients that improve server utilization of such systems over those of strategies known in the current literature.\bigskip

\end{abstract}

\section{Introduction}
\label{sec:intro}



Consider a large distributed computing system that operates over a long time duration. The time duration is modeled as a large number of successive short time slices. The system comprises a number of client computers connected to a number of servers. In each time slice, a client computer can send a unit time job request to a server. Each server is capable of executing exactly one request in each time slice. Hence a server receiving only one request can fulfill the request, while a server requesting multiple requests can fulfill one of the requests and deny the remaining. There is no coordinating agent, and at the beginning of each time slice, the clients sending job requests  each decide on the server to which to make its request independently and simultaneously. In case no one makes a request to a server, that server is unutilized during that time slice. Also if more than one clients send job requests to the same server, all but one will be denied. In this paper, we consider a stylized version of this problem in which (a) the number of clients equal the number of servers and (b) each client sends a job request to a server in each time slice, and devise strategies to minimize the number of denied requests at any time slice after a sufficient number of time slices have elapsed.

This problem is a variant of the Kolkata Paise Restaurant problem \citep[see, e.g.,][]{chakrabarti2009,Ghosh2010,Chakraborti2015}. The restaurant problem is played between a set of $N$ customers and a set of $N$ restaurants over a large number of time periods. In each time period, customers independently and simultaneously choose to visit one among $N$ restaurants. Each restaurant can serve only one customer. If there are more than the one customers choosing a particular restaurant in a particular time period, then the restaurant picks one of these customers at random and serves her, while the other customers choosing that restaurant are not fed. The restaurants are ranked, implying that mere occupation will not keep customers happy as they would try not only to get food, but get it in the best restaurant. 

A related game theoretic formulation is known as the Minority Game problem, which also goes by the name of El Farol Bar problem \citep[see, e.g.,][]{Arthur1994a,Challet2004,Challet1997,Fogel1999}. That problem imagines a bar that can accommodate at most $N_{max}$ people. It is said to become crowded if $N_c$ ($<N_{max}$) or more people visit the bar on any particular evening. There are $N_{max}$ people who decide whether or not to visit the bar on a particular evening. If the bar is crowded, then staying at home is preferable to visiting the bar. However, if the bar is not crowded, then going to the bar is better than staying at home. The game is called a minority game since a player is better off being in the minority.

In our problem, we adopt the feature of the minority game that only a finite number of clients can be served at one location, but the number of locations are exactly equal to the number of clients. Thus every client's objective is to figure out from its own personal history, how to avoid being in majority i.e., clashing with other clients in one location. In the context of client-server mapping, we ignore the ranking scheme of servers. In the literature, it is assumed that information regarding past choices and successful matches of the population are available and clients make use of this information while deciding on the server to choose. In our context, looking up such information at every time slice can be prohibitively expensive, especially since the time slices are short, and hence we restrict the information available to each client to its own history only. Our goal is to maximize the number of fulfilled requests (i.e., the utilization of servers) with simple strategies under this restricted information set.

As a benchmark, it may be noted that if each client chooses each location with identical probability, then the utilization fraction is about 63\% \citep[][ see also Sec.~\ref{sec:lincomb}]{chakrabarti2009} and the maximum utilization rate achieved through rule-of-thumb probabilistic strategies reported in the literature is about 80\% \citep{Ghosh2010}. Most of the strategies considered in the literature often gave rise to significantly smaller utilization rates. 

In this work we provide a number of analytical and numerical results. The first set of results that we state comprises essentially an irrelevance result. We show that the strategies of duplication of other clients' probabilistic strategies or any linear mixing of such strategies across clients (without accounting for information regarding successful matches, either of one's own or of others) lead to a steady state utilization of 63\%, the same as when each client chooses each location with identical probability. This result covers a large set of possible strategies. 

Then we move on to strategies with client-specific history dependence and show that significant improvements in average server utilization are possible using such strategies. We introduce reinforcement learning through a set of strategies, which are differentiated according to the pool of clients who update their respective strategies and the quantum of the update. In all the strategies, clients whose requests are fulfilled at a time slice increase the probability that they assign to sending requests to the same server in future time slices.  A subset of the strategies do not concern themselves with situations in which client requests are denied. Limiting cases in this subset range from ones in which no client updates its probabilities to strategies in which a successful client sticks to the server that fulfilled its request for all future time slices. This set of strategies is a generalization of the Polya urn scheme \citep{Sornette2004}. Other strategies in the set update probabilities for clients whose requests are denied. 

The strategies in the set outperform strategies previously considered in the literature. The intuition is the following. The clients’ objective is to always stay in minority, i.e., to avoid congestion. Under strategies with reinforcement, a successful match increases probability that the successful client will request the same server again and simultaneously the unsuccessful clients reduce their probability of requesting that server. Such an outcome creates an endogenous dispersion of clients over time across servers reducing possible congestion. 

Finally, we study the fitness of these strategies in presence of purely randomizing clients who act as \emph{ noise traders}. The overall utilization fraction decreases unsurprisingly due to their presence. However, it is seen through simulation that, in a relative sense, the clients employing the above strategies outperform noise traders.

This problem can be potentially extended to study the situation where the number of clients is not the same as the number of servers. We primarily stick to a symmetric scenario since in this case, any congestion would automatically imply a simultaneous existence of unused facility, i.e. unused servers as well as unsuccessful clients. If we allow the numbers to differ from each other, then congestion or lack of usage can arise because of mechanical constraints rather than a coordination failure.

The plan of the paper is as follows. In the next section, we state the problem and provide the Pareto and Nash solutions.  In Sec.~\ref{sec:lincomb}, we provide show that large set of probabilistic rules under linear mixing, would converge to uniformly random distribution and hence cannot be used to increase utilization. In Sec.~\ref{sec:update}, we show that some simple rules with updating schemes based on past history significantly improves the utilization ratio. Finally, we introduce heterogeneity in client population i.e. cases where more than one population of clients compete who are following different sets of strategies. We conclude with a summary and discussion.

\section{Pareto efficiency and Nash equilibrium}
\label{sec:pe_ne}
In this section, we first provide a formal description of the game. Then we describe the Nash equilibrium and the Pareto efficient allocation of this game.
Finally, we motivate two classes of na\"{i}ve learning strategies.

\subsection{Description of the game}
The problem that we describe in the introductory section is essentially a game theoretic problem with the following basic structure.
\begin{enumerate}\setlength{\itemsep}{-3pt}
\item Time is discrete.
\item There are $N$ clients and $N$ servers. The set of clients is denoted by $C$ and the set of servers by $R$.
\item At every time slice $t$, each client $i$ sends exactly one connection request to one server.
\item Each server can accommodate at most one client in each time slice. If more than one clients send requests to the server in the same time slice, then one of them is randomly chosen for that time slice. 
\item Payoff of the $i$-th client in time slice $t$ ($\pi_{it}$) is 1 if its request is fulfilled in that time slice, and 0 otherwise.
\item The objective of each client is to secure a match with one server (i.e., maximize $\pi_{it}$) at every time slice.
\end{enumerate}

In the next part of this section, we describe the game theoretic and the most efficient solutions to this problem and show that they are identical \citep{chakrabarti2009}. The Nash equilibrium describes the equilibrium configuration and not the process through which such a configuration might be achieved. It is clear from the nature of the Nash equilibrium  solution that attaining the Nash equilibrium requires tremendous level of coordination among clients. It is unrealistic to assume such a level of coordination is attainable among a large number of competing clients acting independently with only personal information. So in the later portion of the paper, we devise some general rules-of-thumb that are easy to implement and produce some reasonably close results, even though they may not lead to the first best outcome.

\subsection{Pareto efficient allocation and Nash equilibrium}
First, let us define Pareto efficiency and Nash equilibrium. 

\begin{definition}[Allocation]
An `allocation' at time slice $t$ $A_t = \{a_{it}\}_{i \in C}$, where $a_i$ denotes the server to which client $i$ sends a connection request at time slice $t$. Clearly $a_{it} \in R$.
\end{definition}

\begin{definition}[Strategy]
The `strategy' of client $i$ at time slice $t$ is $s_{it}=\{p_{ijt}\}_{j\in R}$, where $p_{ijt}$ is the probability with which client $i$ sends a connection request to server $j$ at time slice $t$. Clearly, $\sum_{j=1}^N p_{ijt} =1$. The strategies of all clients except client $i$ is denoted by $s_{-it}$. The set of all strategies is denoted by $S$.

\end{definition}

\begin{definition}[Pareto efficiency, PE]
A `Pareto efficient' allocation of clients to servers is defined as an allocation for which if one client is reassigned to improve its payoff, then the payoff of at least another client worsens.
\end{definition}
\begin{definition}[Nash equilibrium, NE]
A `Nash equilibrium' is defined as a strategy combination $\{s_{it}^*\}_{i\in C}$ such that for each client $i$ $\pi_i(s_{it}^*)\ge \pi_i(s_{it},s_{-it}^*)$ for all $s_{it} \in S$.
\end{definition}

Clients compete against each other at every time slice. So we can describe the interactions between clients at each time slice as a static one-shot game, which we call a server choice game. For such a game, \cite{chakrabarti2009} and \cite{Mitra_etal;12} show that the set of Pareto efficient outcomes is identical to the set of pure-strategy Nash equilibria through the following proposition.
\begin{proposition}
Consider a server choice game.
Then, allocation $A = \{a_i\}_{i \in C}$ such that $a_i \neq a_j$ for $i \neq j$, $i \in C,\,j \in R$ is PE, and strategy set comprising $s_i = \{p_{ij}\}_{j \in R}$ where $p_{ij} = 1$ if $j = a_i$ and 0 otherwise for all $i \in C$ is NE.  
\label{prop:2_5}
\end{proposition}
{\bf Proof:} \citep[see][]{chakrabarti2009,Mitra_etal;12}. \hfill $\Box$\bigskip

Proposition~\ref{prop:2_5} characterizes the equilibrium configuration, but it does not specify how clients might actually play the game to achieve the PE allocation.
We show below that with some na\"{i}ve learning strategies, we can increase the utilization fraction significantly close to the Pareto allocation. 

We consider two types of na\"{i}ve learning in the next two sections of this paper. In the first type we consider an observation based learning process following \cite{DeGroot_74}. Here, the client strategies are updated based on information about other clients' strategies and not on actual information about past successes and failures. At every time slice, a subset $M \subset C$ of clients is chosen, whose strategies are $s_{mt} = \{p_{mjt}\}_{j\in R}$ for $m\in M$. They update their strategies following the protocol: for $\alpha_i\in [0,1]$, 
\[
p_{mj(t+1)} = \sum_i \alpha_ip_{ijt}, \quad \text{for each }j\in R.
\]
Such a scheme allows copying as well as any arbitrary averaging strategies. We show in Sec.~3 that under some general conditions, such strategies lead to a scenario where every client assigns equal probability to all servers in the steady state.

The second type of na\"{i}ve learning considers updating strategies based on individual successes and failures, i.e.\ clients update their probability of sending a connection request to a server depending on whether their requests were fulfilled or denied in the previous time slice. This type of learning also allows us to consider situations in which the servers do not differentiate between clients,  but the clients believe that there is differentiation, and act accordingly. We consider examples of these strategies in details in Sec.~\ref{sec:update} and show that the average server utilization increases substantially under this type of na\"{i}ve learning.

\section{Probability mixing}
\label{sec:lincomb}
In this section we explore na\"{i}ve learning in which clients update their probabilities by combining their strategies with those of other clients. We show that such mixing does not improve server utilization over scenarios in which clients choose servers at random. We also show that under a general class of observational learning, the clients converge to the same mixed strategy with uniform probability distribution across servers even though they start with arbitrary strategies.

Before analyzing such learning mechanisms, we describe one important result regarding server utilization fraction (i.e., the fraction of the total number of servers that process requests during a time slice; see {\it No learning} strategy in \cite{chakrabarti2009} for a detailed exposition and discussion).

\begin{proposition}[\cite{chakrabarti2009}]
\label{prop:rand}
If all clients assign equal probabilities to all servers, the utilization fraction would be 1-exp(-1).
\end{proposition}
{\bf Proof:} 
We state a general proof here which will be useful later. Suppose there are $N$ servers and $\lambda N$ clients.
Each client assigns $1/ N$  probability to each server. Therefore, the probability that a server is chosen by $x$ clients is given by the binomial distribution
\[
P_x=\binom{\lambda N} {x} \left(\frac{1}{N}\right)^x \left(1-\frac{1}{N}\right)^{\lambda N-x}.
\]
In the limit, the above expression can be approximated by a Poisson distribution,
\[
\lim_{N\rightarrow \infty}P_x=\frac{{\lambda}^x}{x!}\exp(-\lambda)
\]
So the probability that no client chooses one server is 
$P_0=\exp(-\lambda)$ which is $\exp(-1)$ when $\lambda= 1$. This implies that the probability that a server will receive at least one request is $1 - \exp(-1)$.
Note that, numerically $1-\exp(-1) \approx 0.632$. \hfill $\Box$

\subsection{An irrelevance proposition}

We now provide an analytical argument to explain the finding that under a large set of conditions, the average utilization rate is $1-\exp(-1)$.
Below we define two parts of the game and show that if any strategy satisfies the conditions described below, then it will admit a fixed point and we can characterize the fixed point under certain conditions.

Suppose that the $i$-th client's strategy is represented by a row vector $s_i(t)=\{p_{ijt}\}_{j\in R}$. The column vector $S=\{s_1|s_2|\ldots|s_N\}^T$ represents the strategies of all clients. We can model the new strategy vector of any client as the combination of two transformations, a linear transformation and a normalization. 

\paragraph{Stage 1: Linear transformation}~\\
In the first stage, the client chooses to combine any probability assigned by any client to any server. In matrix notation, let us define two column vectors collecting all probability values $p_t=\{p_{11t}, p_{12t},\ldots, p_{1Nt}, p_{21t},$ $ p_{22t},\ldots, p_{N1t}, p_{N2t}, \ldots, p_{NNt}\}^T$ and $q_t=\{q_{11t},q_{12t},\ldots,q_{1Nt},q_{21t},q_{22t},\ldots,q_{N1t},q_{N2t},\ldots,q_{NNt}\}^T$ which we normalize in stage 2. 
Let us also define a weight matrix $W$ with non-negative entries. The linear transformation step of the strategy mixing process computes $q_t$ as
\[ q_t = W \cdot p_t.
\]


Thus, for a generic $(i,j)$ pair, 
\begin{equation}
q_{ijt}=\sum_{k=1}^N \sum_{l=1}^N w^{ij}_{kl} p_{klt}.
\label{Eqn:lincomb}
\end{equation}

Note that here we do not impose any averaging condition i.e., row sums are not necessarily equal to 1. We will impose such a condition in the next stage.

\paragraph{Stage 2: Normalization}~\\
Since $\sum_{j=1}^N q_{ijt} \ne 1$ at time slice $t$, for any $i\in N$ in general, $q_{ijt}$ values cannot be treated as probabilities.
So we normalize them as follows. We define a diagonal matrix $Q_{it} = (1/\sum_{j=0}^N q_{ijt})I$ where $I$ is an identity matrix of size $N$. 
Then the components of the vector $Q_{it} \cdot q_{it}$ can be treated as probabilities.
Using this notation, we can write 
\[
p_{(t+1)} = Q_t \cdot q_t,
\] 
where $Q_t$ is a block diagonal matrix whose $i$-th diagonal component is the matrix $Q_{it}$, and $p_{ij(t+1)}$ in terms of $p_{ijt}$ values as 
\begin{equation}
p_{ij(t+1)}=\frac{\sum_{k=1}^N \sum_{l=1}^N w^{ij}_{kl} p_{klt}}{\sum_{j=1}^N \sum_{k=1}^N \sum_{l=1}^N w^{ij}_{kl} p_{klt}}.
\label{Eqn:p_evolve}
\end{equation}

%


\subsection{Existence of a fixed point}
Let us now restrict the discussion in the previous section to a strategy vector for all clients.
The strategy vector belongs to a simplex
\[
\sum_{i=1}^N \sum_{j=1}^N p_{ij}=N.
\]

Therefore, defining 
\[
\Delta^{N^2} =\{ p \in \mathbb{R}^{N^2}\bigg{|} \sum_{i=1}^N \sum_{j=1}^N p_{ij}=N\}.
\]
the whole updating scheme can be written as a mapping 
\[
f: \Delta^{N^2}\rightarrow \Delta^{N^2}
\]
Note that $f$ is continuous in $p$, and that $\Delta^{N^2}$ is a convex and compact subset of the Euclidean space. 
To prove existence of a fixed point of the mapping, we use the following theorem \citep[see, e.g.][]{Carter2001}. 
\begin{theorem}[Brower's fixed point theorem]
Every continuous function from a convex compact subset of a Euclidean space to itself has a fixed point.
\end{theorem}

\begin{proposition}
A necessary condition for assigning uniform probability to each server by each clients is that 
\[\sum_{k=1}^N \sum_{l=1}^N w^{ij}_{kl} \text{ is a constant.}
\]
\end{proposition}

\noindent {\bf Proof:}
Suppose $p_{ijt}=1/N$. Then from Eqn.~\ref{Eqn:p_evolve},
\[
p_{ij(t+1)} =\frac{\sum_{k=1}^N \sum_{l=1}^N w^{ij}_{kl} p_{klt}}{\sum_{j=1}^N \sum_{k=1}^N \sum_{l=1}^N w^{ij}_{kl} p_{klt}} 
=\frac{\sum_{k=1}^N \sum_{l=1}^N w^{ij}_{kl}}{\sum_{j=1}^N \sum_{k=1}^N \sum_{l=1}^N w^{ij}_{kl}}
\]
which is guaranteed to be $1/N$ only if $\sum_{k=1}^N \sum_{l=1}^N w^{ij}_{kl}$ is a constant.\hfill$\Box$

\subsection{Convergence to uniform probability}

Let us now study convergence to uniform probability distribution. For this purpose, we assume the existence of mechanisms such that
\begin{equation}
\sum_{j=1}^N q_{ijt}=1 \quad\text{for all }i\in C.
\label{Eqn:q_sum}
\end{equation}
Prop.~\ref{Prop:3.4} describes a necessary condition for the existence of such a strategy. 
\begin{proposition}
A condition necessary to treat the column vector $q$ as the updated strategy vector $p_{(t+1)}$ is
\begin{equation}
\left(\sum_{j=1}^N w^{ij}_{11}\right) p_{11t}+\left(\sum_{j=1}^N w^{ij}_{12}\right) p_{12t}+\ldots+\left(\sum_{j=1}^N w^{ij}_{NN}\right) p_{NNt}=1,
\label{Eqn:q_cond}
\end{equation}
for all $i\in N$.
\label{Prop:3.4}
\end{proposition}
{\bf Proof:} Eqn.~\ref{Eqn:q_sum} can be rewritten as
\begin{equation}
\sum_{j=1}^N\sum_{k=1}^N \sum_{l=1}^N w^{ij}_{kl} p_{klt}=1,
\end{equation} 
which in turn can be expressed as Eqn.~\ref{Eqn:q_cond} using Eqn.~\ref{Eqn:lincomb}.\hfill$\Box$

\begin{corollary}
A necessary condition for uniform probability is
\begin{equation}
\left(\sum_{j=1}^N w^{ij}_{11}\right) +\left(\sum_{j=1}^N w^{ij}_{12}\right)+ \ldots+\left(\sum_{j=1}^N w^{ij}_{NN}\right)=N,
\end{equation}
\end{corollary}
{\bf Proof:}
Follows directly from Eqn. \ref{Eqn:q_cond} by substituting $p_{klt}=1/N$ for all $k,l\in N$ and $t$.\hfill$\Box$\bigskip

Under conditions satisfying Prop.~\ref{Prop:3.4}, the second part of the transformation i.e. normalization, is redundant and 
\begin{equation}
p(t+1) = W \cdot p(t).
\label{Eqn:p_dynamic}
\end{equation}\bigskip


We now describe conditions under which Eqn.~\ref{Eqn:q_cond} describes a strategy evolution process by characterizing two steps. 
In the first step, we show the convergence result for Eqn.~\ref{Eqn:p_dynamic} for generic matrix $W$. In the second step, we provide conditions for all clients assign uniform probability to all servers. The first step relies on Markov chain theory and the second step relies on consensus formation in DeGroot model \citep{Jackson_network_10}. We work under the following assumption.

\begin{assumption} The row sum of the weight matrix is 1 i.e. for each client $i$, each server $j$, $\sum_{k=1}^N\sum_{l=1}^N w^{ij}_{kl}=1$.
\end{assumption}

We will use the following definitions related to matrices.

\begin{definition}[Strongly connected matrix]
An $N \times N$ square matrix $W$ is `strongly connected' if for each pair $(i,j) \in N \times N$ there exists a sequence of distinct indices $i = r_0, r_1, \dots, r_n = j$ such that $w_{r_kr_{(k+1)}} > 0$ for $k = 0, \dots, n-1$.
\end{definition}


\begin{definition}[Directed cycle and cycle length]
Consider an $N \times N$ square matrix $W$ and a sequence of distinct indices $i = r_0, r_1, \dots, r_n$ in $W$ such that $w_{r_kr_{(k+1)}} > 0$ for $k = 0, \dots, n-1$ and $w_{r_n r_0} > 0$. Such a sequence is called a `directed cycle' of `length' $n$.
\end{definition}

\begin{definition}[Aperiodic matrix]
An $N \times N$ square matrix $A$ is `aperiodic' if the lengths of all its directed cycles are co-primes.
\end{definition}

Prop.~\ref{prop:conv} below shows that convergence occurs. 
\begin{proposition}
The strategy vector $P$ converges if and only if every set of indices in $W$ that is strongly connected and closed is aperiodic.
\label{prop:conv}
\end{proposition}
{\bf Proof:} This is a standard result based on Markov chains \citep[see, e.g.,][]{Jackson_network_10}.\hfill$\Box$\bigskip

Next, we show that each of the clients assign uniform probability.

\begin{proposition}[\cite{Jackson_network_10}]
Under the updating matrix $W$, any strongly connected and closed group of individuals
assigns probabilities $1/N$ for every initial strategies if and only if it is aperiodic.
\end{proposition}
{\bf Proof:} This proposition depends on two statements. Proposition 8.3.1 in \cite{Jackson_network_10} provides the necessary and sufficient conditions
for convergence to consensus where every client assigns same probability.
Since for each client, the probabilities must sum up to one, each probability must be $1/N$.

Therefore, all clients will assign uniform probability to all servers, i.e., $
p_{ijt}|_{t\rightarrow \infty}=\frac{1}{N}$.\hfill$\Box$


\section{Updating strategies}
\label{sec:update}
We have so far considered cases in which agents do not ``learn'' from previous successes or failures when updating their strategies but modify their strategies by observing those of other clients. We have seen that for a large class of such strategies, only about 63.2\% of the servers fulfil requests at each time slice in the steady state when the number of servers is large (see Sec. \ref{sec:lincomb}). The present section is therefore devoted to strategies  based on a client's own past performance in order to increase the utilization fraction of the client server system. Under these strategies, clients do not use any information about other clients' performances or about past server utilization while updating strategies. They only retain their own history of successes/failures and attempts.

As before, we use the utilization fraction of servers, defined as the fraction of the total number of servers that fulfil client requests during that time slice to measure the performance of a strategy. We also use another measure called the stability fraction. It is the fraction of clients that assign a high enough probability for selecting one particular server. We used a probability threshold value of 0.99. 

For each of the strategies described in this section, every client starts by assigning equal probability $1/N$ of sending connection requests to each server. 

\subsection{Updating by successful clients only}
\label{subsec:success}

Consider the following update strategy. Suppose a client $i$ sent a request to server $r$ in time slice $t$. If its request was fulfilled by the server, then it will opt for that particular server in all future time slices. If however, the request was denied by the server, then in time slice $t+1$, then client $i$ does not update its probability distribution when selecting servers in time slice $t+1$.

In other words, if client $i$'s request has been fulfilled by server $r$ in time slice $t$, then for all time slices $\tau > t$
\begin{equation}
p_{ij\tau} = \left\{\begin{array}{ll}
1 & \text{if } j=r;\\
0 &\text{otherwise.}
\end{array}\right.
\label{strat1_succ}
\end{equation}
On the other hand if the request was denied, then 
\[p_{ij(t+1)} = p_{ijt} \text{ for all }j \in R.\]
We refer to this client probability updating strategy as STRATEGY 1.

We see from (\ref{strat1_succ}) that any client whose request has been served at any point becomes a stable client, so that the stability fraction will increase to a value of 1 as the number of time slices increase. The utilization fraction also increases and stabilizes at a value close to 0.8. It does not reach a value of 1 since more than one clients can become stable after their requests are fulfilled by the same server in different time slices. 

\begin{proposition}
\label{prop:update_success}
If the clients follow STRATEGY 1 then the stability fraction tends to 1 and utilization fraction is in the interval [0.79, 0.81].
\end{proposition}
{\bf Proof :} We can show it by using Prop.~\ref{prop:rand}. See App.~\ref{sec:app} for a detailed proof.\hfill$\Box$\bigskip

\begin{figure}[!hbt]
\centering
\begin{tabular}{cc}
\includegraphics[width=3.1in]{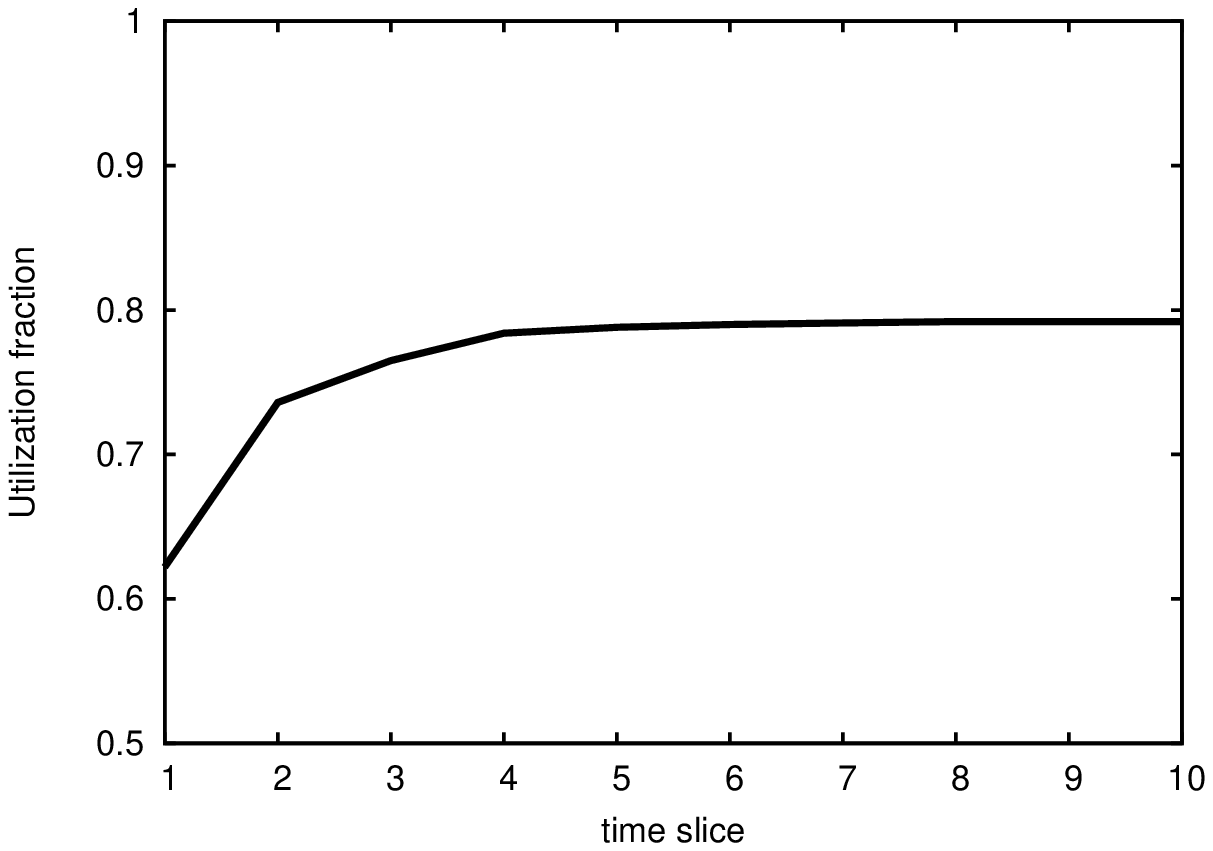} 
&
\includegraphics[width=3.1in]{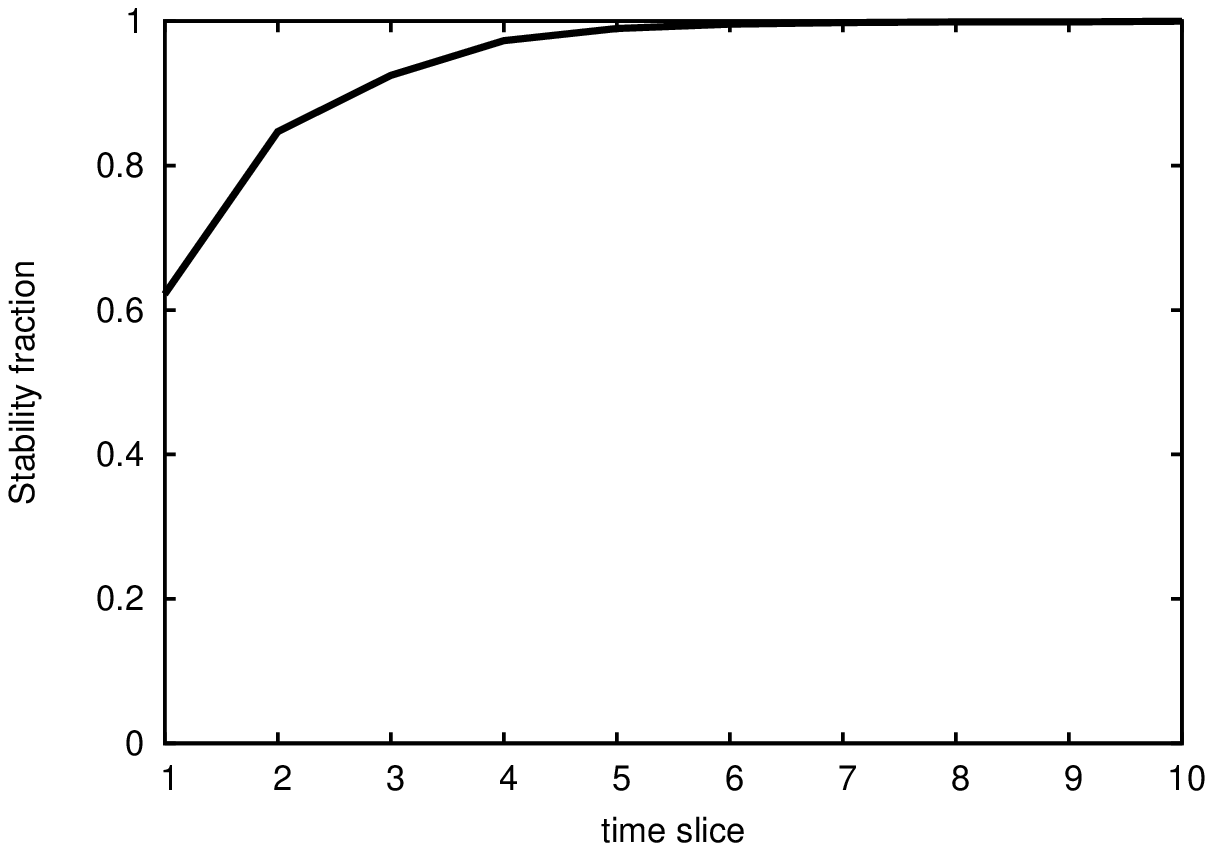}\\[30pt]
Utilization fraction & Stability fraction
\end{tabular}
\caption{Variation of utilization fraction and stability fraction for the set of clients following a non-stochastic updating strategy}
\label{fig:ss1}
\end{figure}

Figure~\ref{fig:ss1} shows the variation in utilization fraction and stability fraction for a set of 1000 clients following STRATEGY~1 requesting 1000 servers over 10 time slices. Note that the utilization fraction plotted for a time slice is computed during each time slice, while the stability fraction plotted for a time slice is computed after requests have been processed during that time slice. From this experiment we see that the utilization fraction is approximately 0.8, which is within the range stated in Prop.~\ref{prop:update_success}.

\subsection{Updating by all clients}
\label{subsec:all}
Starting with client probability distributions identical to those at the start of STRATEGY 1, we now present strategies that result in utilization fractions higher than those from STRATEGY 1 in the long term, i.e., after a large number of time slices. In these strategies, called STRATEGY 2A and STRATEGY 2B, clients whose requests have been denied during a time slice also update their probability distributions in the next time slice.

\paragraph{STRATEGY 2A:} 
Consider a client $i$ that has sent a request to server $r$ in time slice $t$. If the request was fulfilled, then $i$ increases the probability of requesting $r$ in time slice $t+1$ by a constant amount $s$ or to 1 if that is not possible. This increase is compensated by a proportional decrease in the probabilities of requesting other servers. This means that if $p_{irt} < 1$ then 
\begin{equation}
p_{ij(t+1)} = \left\{\begin{array}{ll}
p_{ijt} + \delta & \text{if } j=r;\\
(1 - \alpha)p_{ijt} &\text{otherwise;}
\end{array}\right.
\label{strat2a_succ}
\end{equation}
where $\delta = \min\{s, 1 - p_{irt}\}$ and $\alpha = \delta/(1-p_{irt})$. If $p_{irt} = 1$, then $p_{ij(t+1)} = p_{ijt}$ for all $j \in R$.

If the client's request has been denied in time slice $t$, then it reduces the probability of requesting server $r$ in time slice $t+1$ by a constant amount $s$ or to 0 if that is not possible. In case $p_{irt} < 1$, this decrease is compensated by a proportional increase in the probabilities of requesting other servers. Thus
\begin{equation}
p_{im(t+1)} = \left\{\begin{array}{ll}
p_{imt} - \theta & \text{if } m=r;\\
(1 + \beta)p_{imt} &\text{otherwise;}
\end{array}\right.
\label{strat2a_fail.1}
\end{equation}
where $\theta = \min\{s, p_{irt}\}$ and $\beta = \theta/(1-p_{irt})$.

If $p_{irt} = 1$ in time slice $t$, then such a  proportionate increase is not possible. In this situation the increase is equal for all servers, i.e.,  
\begin{equation}
p_{im(t+1)} = \left\{\begin{array}{ll}
p_{imt} - \theta & \text{if } m=r;\\
\theta/(N-1) & \text{otherwise.}
\end{array}\right.
\label{strat2a_fail.2}
\end{equation}

\paragraph{STRATEGY 2B:} 
This strategy is a variation of STRATEGY 2A. In STRATEGY 2A, if client $i$ sends a request to server $r$ in time slice $t$, the value of $p_{ir(t+1)}$ usually differs from the $p_{irt}$ value by a constant amount if not naturally truncated. In STRATEGY 2B, if client $i$'s request in time slice $t$ has been fulfilled by server $r$, then $p_{irt}$ increases by a constant fraction $f$ of the amount $(1 - p_{irt})$. On the other hand if the request was denied in time slice $t$, then $p_{irt}$ reduces by the same fraction $f$ of the $p_{irt}$ value. Hence for this strategy, the form of the expressions (\ref{strat2a_succ}) through (\ref{strat2a_fail.2}) remain unchanged, but the values of $\delta$ and $\theta$ change to 
\begin{align*}
\delta & = f(1 - p_{irt})\text{ and }\\
\theta & = f\, p_{irt}.
\end{align*}
Notice that following this strategy, the probability that a particular client requests a particular server at any time slice will never be either 0 or 1 if it was initially not set to either 0 or 1 (unless of course, $f = 1$).

\begin{figure}[!hbt]
\centering
\begin{tabular}{cc}
\includegraphics[width=3.1in]{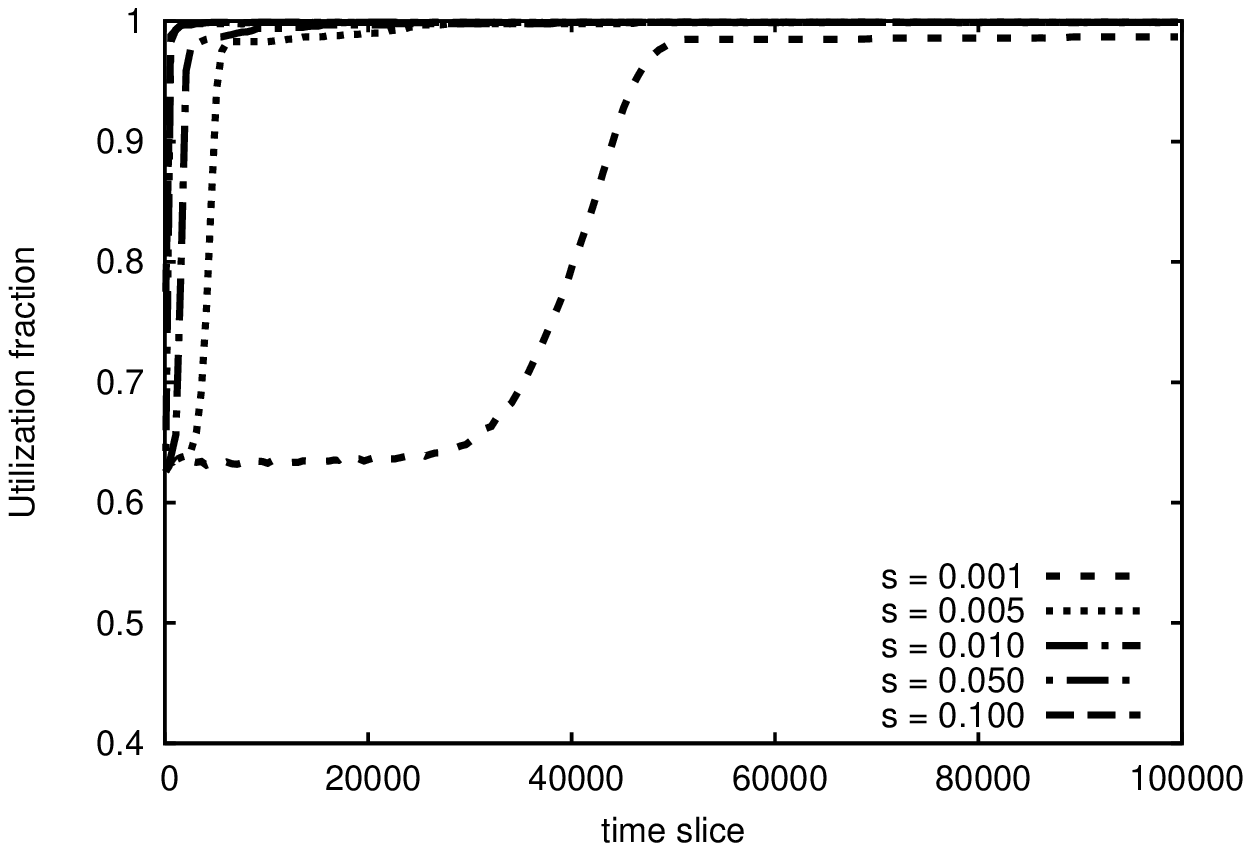} &
\includegraphics[width=3.1in]{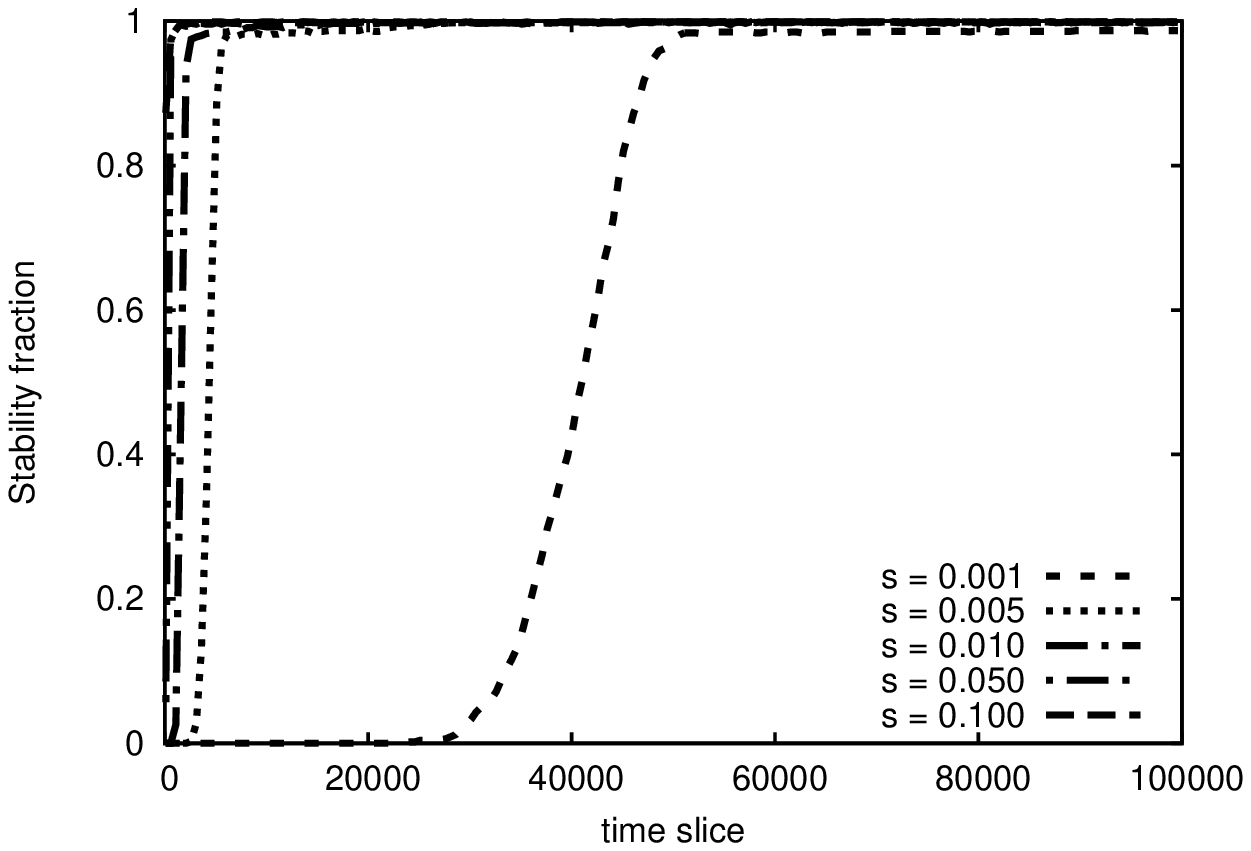} 
\end{tabular}\\[30pt]
\caption{Variation of utilization and stability fractions with rounds when customers follow STRATEGY 2A}
\label{fig:ss2a}
\end{figure}

\begin{figure}[!hbt]
\centering
\begin{tabular}{cc}
\includegraphics[width=3.1in]{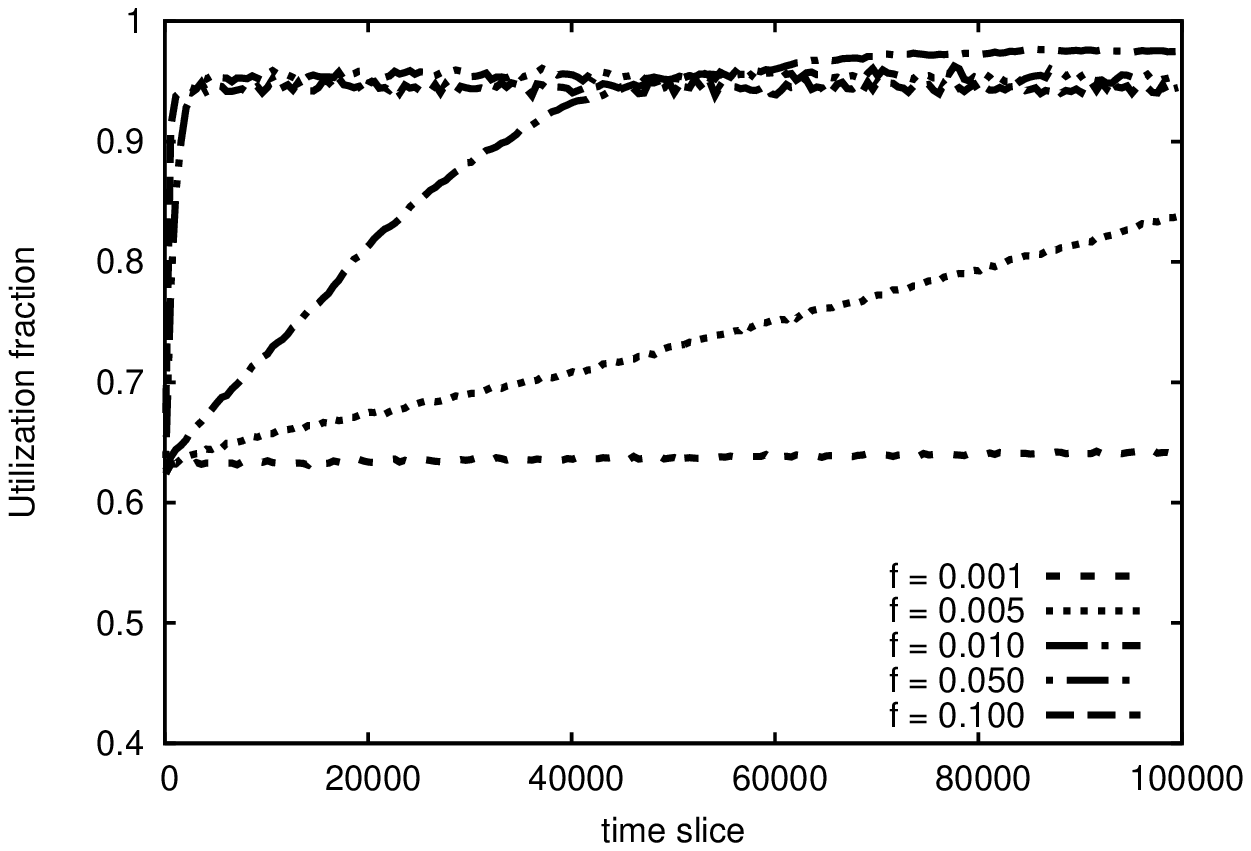} &
\includegraphics[width=3.1in]{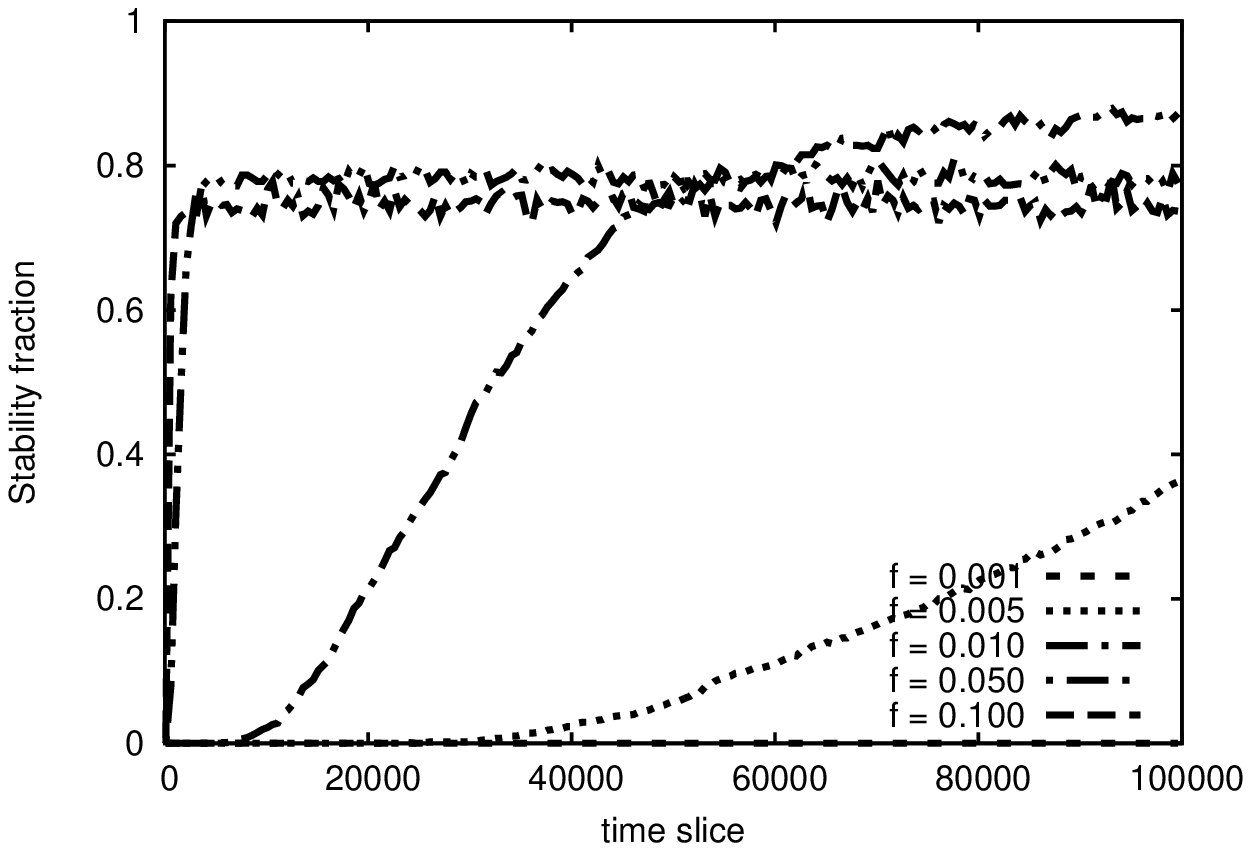} 
\end{tabular}\\[30pt]
\caption{Variation of utilization and stability fractions with rounds when customers follow STRATEGY 2B}
\label{fig:ss2b}
\end{figure}

Figure~\ref{fig:ss2a} shows the variation in utilization fraction and stability fraction with time slices if customers follow STRATEGY 2A. The value of $N$ is taken as 1000 and the step size $s$ is varied from 0.001 to 0.1 in the plots. 
Figure~\ref{fig:ss2b} shows the variation in utilization fraction and stability fraction with time slices if customers follow STRATEGY 2B. The value of $N$ is taken as 1000 and  the fraction $f$ is varied from 0.001 to 0.1 in the plots. 

\subsection{Updating based on perceived history dependence}
\label{subsec:Polya}
Next we consider a client strategy in which clients choose to connect to servers that have fulfilled their requests in previous time slices. Hence clients behave as if servers that have favored them in the past actually preferred to serve them over other clients. We refer to this strategy as STRATEGY POLYA. Servers with multiple request of course choose clients at random, hence this is just a ``perceived'' dependence on historical fulfillment of requests from the clients' perspective.

This strategy is in line with the Polya's urn model \citep{Sornette2004} that allows us to introduce reinforcement learning for server choice. The essential idea is that if one request is fulfilled, then the client reinforces choice of the same in the updated version. In one limit, no updating occurs. In the other end, extreme updating occurs where a client assigns the full probability to one server that served the client just once. Note that this allows us to interpolate between completely random strategies and STRATEGY 1.

Consider $N$ clients and $N$ servers. Clients update strategies as follows. Prior to the first round every client allots a value of 1 to each server. Then it normalizes the values to obtain a probability $1/N$ of choosing each of the $N$ servers. If a client now sends a request to server $r$ in the first time slice and this request is fulfilled, she increases the value assigned to that server by an amount $k = mN/(N-m)$ where $m$ is a multiplier that is input to the process. When $N$ is large, $m = 1$ will mean that the value assigned to the server $r$ is increased by approximately 1, leading to a very small increase in the preference for the server $r$, while a value of $m$ close to $N$ implies that server $r$ would definitely be chosen by client $i$ in future time slices. 

Therefore at the end of time slice $t$, if the number of times client $i$'s connection request has been fulfilled by the $j$-th server $n_{ijt}$ times, then the probability of client $i$ choosing server $j$ in time slice $(t+1)$ is  
\begin{equation}
p_{ij(t+1)} = \frac{1+k\,n_{ijt}}{N + k\sum_{l=1}^N n_{ilt}}.
\end{equation} 

Note that the effect of these modifications on weights diminish as the number of time slices increase. Hence the utilization fraction will stabilise asymptotically if clients follow this strategy. Also, when $N$ becomes large, the changes in probabilities will be small for small values of $m$, and the strategy will be very similar to the one in which a customer chooses a restaurant at random with equal probabilities. 

\begin{figure}[!hbt]
\centering
\includegraphics[width=3.1in]{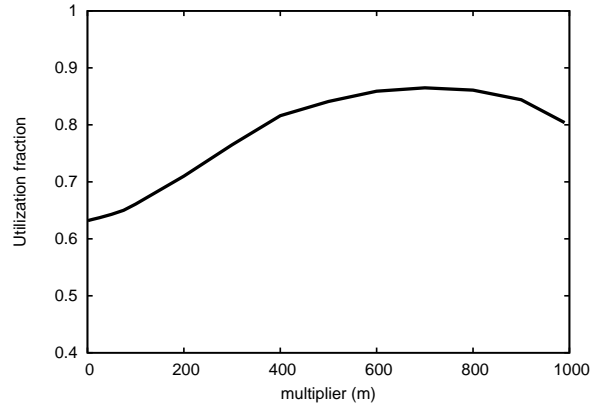}\\[30pt]
\caption{Variation in the utilization fraction with increasing values of the multiplier $m$}
\label{fig:polya}
\end{figure}

We observed the values of the utilization fractions at the end of 10000 time slices when 1000 clients following STRATEGY POLYA sent requests to 1000 servers over 10000 time slices, and the value of the multiplier $m$ was varied from 1 through 1000. The result is shown in Figure~\ref{fig:polya}.

These results have interesting connections with other strategies. When $m = 0$, $k$ becomes 0 and the $p_{ijt}$ values all remain $1/N$, i.e., there is no updating. This is identical to the situation described in Prop.~\ref{prop:rand} and the utilization fraction is approximately 0.632. When $m = 1$, $k$ is approximately 1 for large $N$ and this corresponds to Polya's urn model \citep[see e.g.,][]{Sornette2004}. When $m$ is sufficiently large, $k$ tends to $\infty$; hence STRATEGY POLYA converges to STRATEGY 1 and we see the utilization fraction is approximately 0.8.

\section{Heterogeneity in client strategies}
\label{sec:mixing} 
In this section we combine a set of clients following one of the strategies presented in Section~\ref{sec:update} with a set of clients who choose servers at random. We examine the effect of this mixing on the overall utilization fraction of all clients, and that of the set of clients following a strategy from Section~\ref{sec:update}, and show that clients who update strategies based on their own past successes are better off than the rest even in mixed populations.

Consider the set $S$ of clients partitioned into two sets $S_s$ and $S_r$. Clients in $S_s$ update probabilities of choosing servers to make requests using one of the strategies described in Section~\ref{sec:update}, while clients in $S_r$ choose servers to make requests at random, i.e., they do not update their probabilities of choosing servers. 

\begin{figure}[!hbt]
\centering
\includegraphics[width=3.1in]{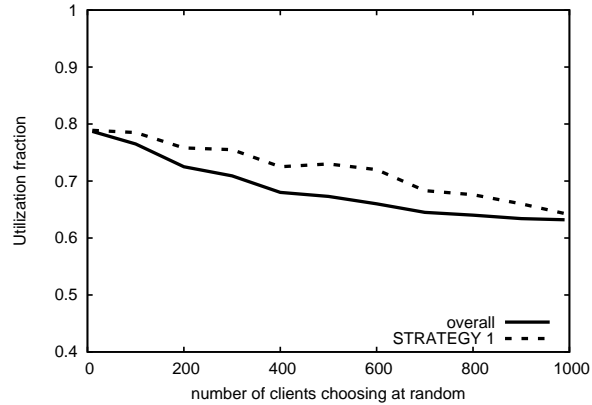}\\[30pt]
\caption{Effect of mixing clients following STRATEGY 1 with clients choosing servers at random on utilization fraction (total number of clients = 1000)}
\label{fig:strat1mix}
\end{figure}

\begin{figure}[!hbt]
\centering
\begin{tabular}{cc}
\includegraphics[width=3.1in]{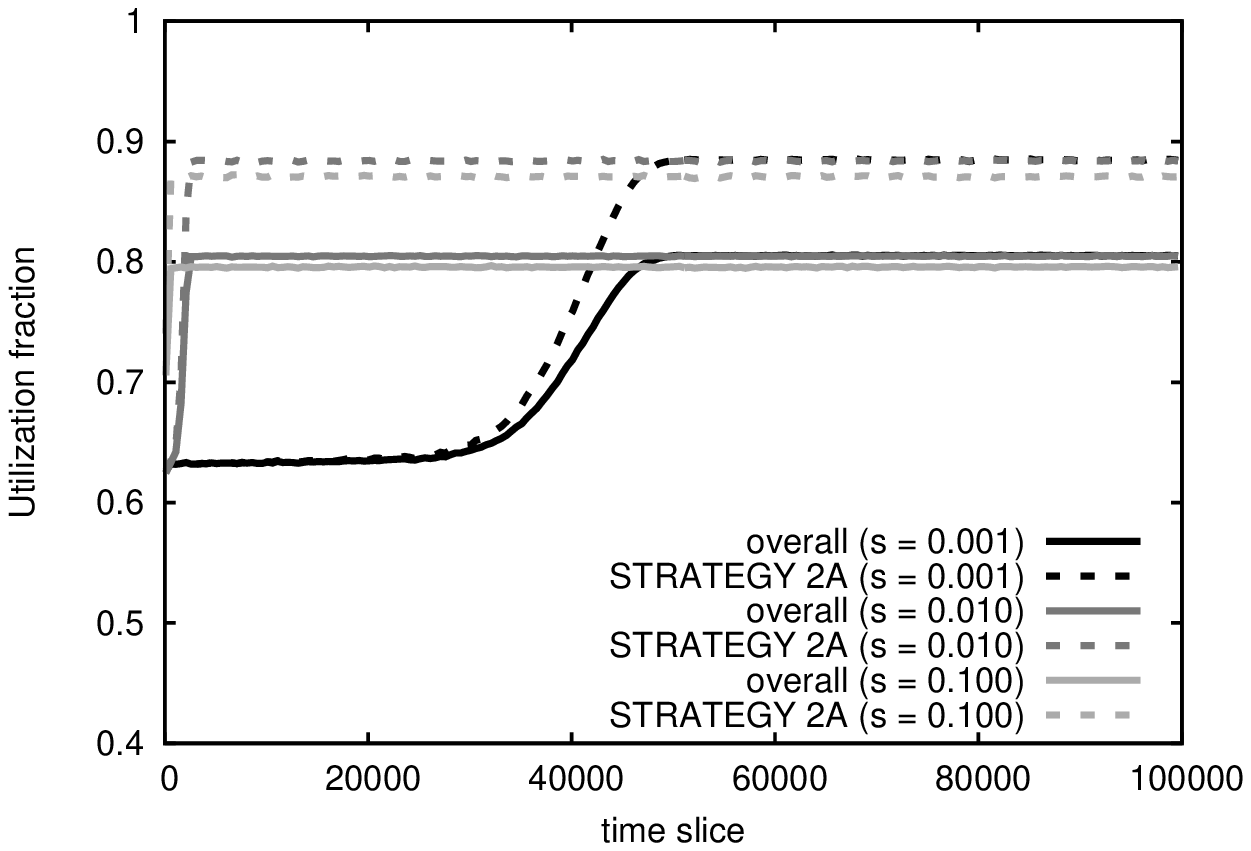} &
\includegraphics[width=3.1in]{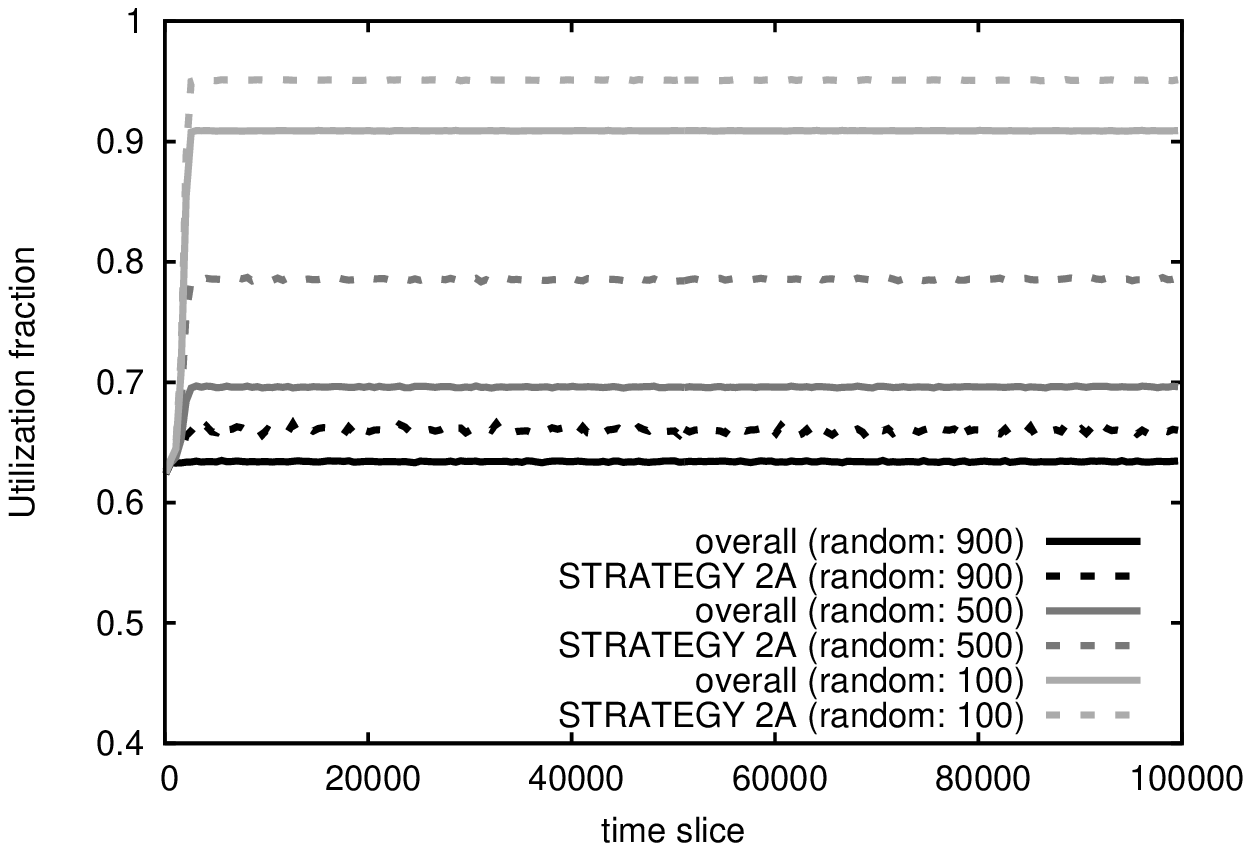}
\end{tabular}\\[30pt]

\caption{Effect of mixing clients following STRATEGY 2A with clients choosing servers at random on utilization fraction (total number of clients = 1000)}
\label{fig:strat2Amix}
\end{figure}

\begin{figure}
\centering
\begin{tabular}{cc}
\includegraphics[width=3.1in]{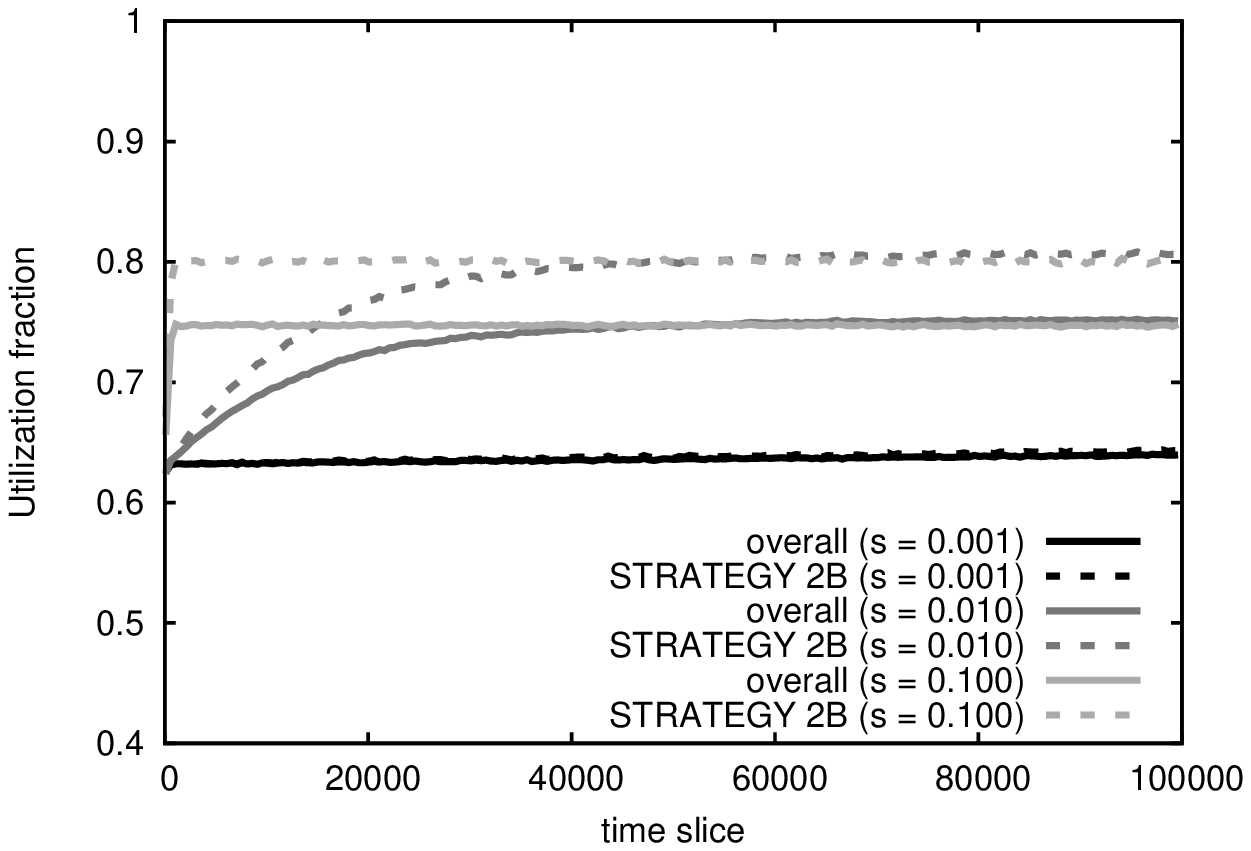} &
\includegraphics[width=3.1in]{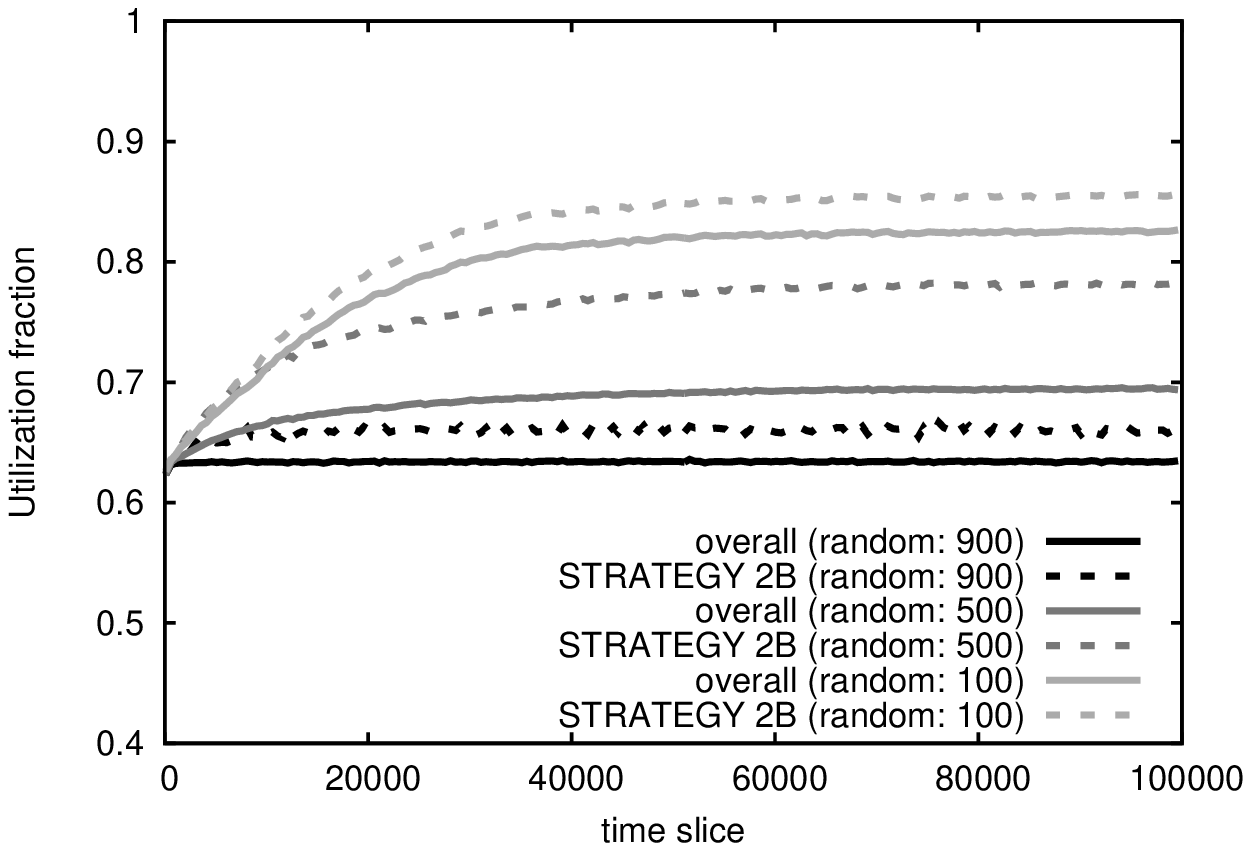}
\end{tabular}\\[30pt]

\caption{Effect of mixing clients following STRATEGY 2B with clients choosing servers at random on utilization fraction (total number of clients = 1000)}
\label{fig:strat2Bmix}
\end{figure}

\begin{figure}
\centering
\includegraphics[width=3.1in]{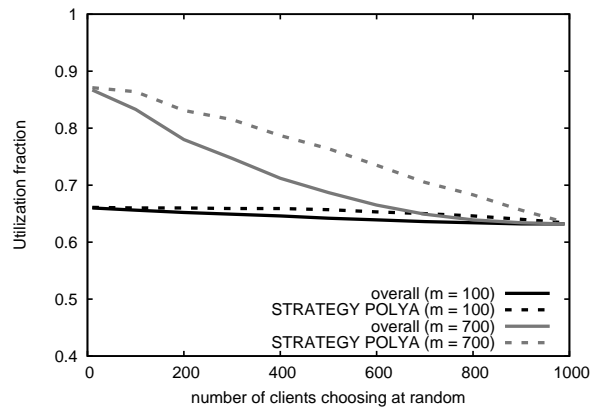}\\[30pt]
\caption{Effect of mixing clients following STRATEGY POLYA with clients choosing servers at random on utilization fraction (total number of clients = 1000)}
\label{fig:polyamix}
\end{figure}

If clients in $S_s$ follow STRATEGY 1, the utilization fractions of clients in $S_s$ and $S_r$ converge to their stable values within the first few iterations. Hence in this case we only look at the stable values of the utilization fractions. Figure~\ref{fig:strat1mix} shows the variation in these utilization fractions of the set of all clients and of clients in $S_s$ when clients in $S_s$ update probabilities using STRATEGY 1 and $(|S_s|, |S_r|)$ varies from (10, 990) to (990, 10).
The plot on the left hand side of Figure~\ref{fig:strat2Amix} shows the variation in utilization fractions of all clients and clients in $S_s$ when the mix of clients is kept constant at $(|S_s|, |S_r|) = (750, 250)$, clients in $S_s$ follow STRATEGY 2A, and the value of $s$ is varied between 0.001 and 0.01. The plot on the right hand side shows this variation when the value of $s$ is kept constant at 0.01 and the mix $(|S-s|, |S_r|)$ changes from (900, 100) to (100, 900). Figure~\ref{fig:strat2Bmix} presents results from similar experiments when clients in $S_s$ follow STRATEGY 2B.
If clients in $S-s$ follow STRATEGY POLYA, the utilization fractions of clients in $S_s$ and $S_r$ stabilize in the first few time slices. (This is similar to the situation when they follow STRATEGY 1.) Figure~\ref{fig:polyamix} shows the variation in the stabilized utilization fractions of the set of all clients and of clients in $S_s$ when clients in $S_s$ update probabilities using STRATEGY POLYA and $(|S_s|, |S_r|)$ varies from (10, 990) to (990, 10).

The results from mixing a set of clients that follow an updating strategy from Section~\ref{sec:update} with clients who do not update their probabilities follow a couple of common trends. First, the average utilization fractions of the whole set of clients decrease as the proportion of clients that do not update probabilities increase in the population of clients. This decrease is steepest when the average utilization fractions increase the most as a result of clients following a particular strategy with a particular set of parameters. Second, in all the mixes considered, the average utilization values of the set of clients that update probabilities is higher than that of the whole population of clients. Hence it makes sense for a client to adopt an updating strategy even if other clients in the population do not do so.

\section{Summary}
\label{sec:summary}
In this paper, we have considered a distributed computing system in which there are $N$ servers and $N$ clients interacting over multiple time slices. The clients make job requests to servers at every time slice. The servers can fulfil only one request in a time slice, and so if more than one clients request the same server, all but one randomly chosen client's request are denied. There is no coordinating agent and the clients decide on which server to send job requests simultaneously and independently. We have modeled this as a game in which there are $N$ servers and $N$ clients, in which a server remaining unutilized in a time slice implies that a client request is denied in that time slice at some other server. This game essentially captures the idea of coordination failures in a multi-agent environment. In this regard, it draws parallels with the Kolkata Paise Restaurant problem and minority games. (See Sec.~\ref{sec:intro} for details.)


This paper makes two contributions. One, it provides a set of sufficient conditions such that if some updating strategies satisfy them, the game would converge to one in which players make their choices at random. Hence, it works as an irrelevance result stating that considering any such strategies is effectively useless. (See Sec.~\ref{sec:lincomb} for these results.) Secondly and more importantly, we propose strategies that use only clients' own personal information and performs much better than the previously proposed strategies e.g., in \cite{Chakraborti2015} in terms of increasing utilization fraction. (See Sec.~\ref{sec:update} for these strategies.)

We introduce the idea of reinforcement learning through a generalized scheme of Polya's urn model. It has two limits. The first limit is a random choice scheme which indicates zero reinforcement producing about 63\% utilization fraction. In the other end with extreme reinforcement, we show the the utilization fraction is approximately 80\%. (These results also appear in Sec.~\ref{sec:update}.)

Further experiments with heterogeneous client population shows that
although presence of noise traders in the form of clients who make choices purely at random reduce the overall server utilization, clients with the suggested strategies outperform them. Thus employing such strategies are better even with noise traders. (See the discussion in Sec.~\ref{sec:mixing}.) 

The work reported here can be useful for maximizing global efficiency of massive systems with parts operating in parallel, that need to secure connections between complementary parts repeatedly. Rather than using protocols that require collection of information across the system, such simple modes of distributed computation increase system efficiency at low implementation costs.

\bigskip

\noindent{\bf Acknowledgement:} The first author is grateful to Bikas K. Chakrabarti for some useful discussions.

\section{Appendix}
\label{sec:app}

{\bf Proof of Prop. \ref{prop:update_success} :} Recall from Prop. \ref{prop:rand} that with $N$ servers and $\lambda N$ clients, the fraction of utilized servers would be
\[
f=1-exp(-\lambda).
\label{Eqn:frac}
\]

\noindent Let us consider this case time slice by time slice. See Fig. \ref{fig:ss1} for comparing 
the values of the utilization fraction and the
stability fraction we derive in the following with the simulation results.
\begin{itemize}
\item[$t$= 1 :] In the first time slice, there are $N$ clients and $N$ servers, with each client assigning $1/N$ probability to each server. Therefore, by applying Eqn. \ref{Eqn:frac} from Prop.
\ref{prop:rand} that the first time slice's utilization fraction is approximately
\[ 
f_1 \approx 0.632.
\]
Following the same calculations, the stabilization fraction in the first time slice is
\[ 
\theta_1 \approx 0.632.
\]

\item[$t$= 2 :] In the next time slice, $f_1$ fraction of clients have stabilized following their strategy protocol. Around 0.368 fraction could not secure a server and hence, have not stabilized either.
In this time slice, they will choose randomly. So $(1-f_1)N$ number of clients would choose randomly from $N$ servers. Again by applying Eqn. \ref{Eqn:frac} , we see that the new fraction of
potentially successful matches would be
\begin{eqnarray*} 
f^{pot}_2 &\approx & 1-exp(-0.368) \nonumber \\
                          &\approx & 0.307. 
\end{eqnarray*}
However, there will be out of these many new potential matches, there will be crowding in the servers already chosen by those already stabilized in the first time slice. Therefore,
\begin{eqnarray*} 
f_2 &\approx & 0.632+0.368 \times 0.307 \nonumber \\
                          &\approx & 0.744. 
\end{eqnarray*}
Computing the stabilization fraction is a bit more involved. Out of the new $0.307$ fraction, $0.632 \times 0.307$ fraction do crowd into servers that were utilized in the first time slice.
Since the servers do not differentiate across clients, the fraction of stabilized clients (old clients+new clients conflicting with old clients + new clients in new locations) would be
\begin{eqnarray*} 
\theta_2 &\approx & 0.632+ (0.632\times0.307)/2+ 0.368 \times 0.307 \nonumber \\
                          &\approx & 0.842. 
\end{eqnarray*}
\item[$t$= 3 :] Thus the fraction of unstabilized clients are now $1-\theta_2$ = 0.158.
Therefore, following same logic,
\begin{eqnarray*} 
f_3 &\approx & 0.744+ (1-exp(-0.158))\times (1-0.744) \nonumber \\
                          &\approx & 0.781. 
\end{eqnarray*}
To compute the stabilization fraction, let us ignore the possibility that there might be more than 2 clients at one server, for the sake of tractability.
Therefore, the new fraction of stabilized clients 
(old clients+new clients conflicting with old clients + new clients in new locations) would be
\begin{eqnarray*} 
\theta_3 &\approx & 0.842+ (0.744\times (1-exp(-0.158))/2+ (1-0.744) \times (1-exp(-0.158)) \nonumber \\
                          &\approx &  0.933.
\end{eqnarray*}
\end{itemize}
\noindent One can go on computing in a similar fashion. For $t=4$, $\theta_4$ would be approximately 0.98 and $f_4$ would be approximately 0.79.

As can be verified from Fig \ref{fig:ss1}, by the fourth time slice about 98\% of clients would have stabilized utilizing more than 79\% servers. 
Simulation shows that the subsequent gains in utilization fraction is very small as it converges as the population stabilizes quite fast (see Fig. \ref{fig:ss1}). 
Therefore the range of the limiting utilization fraction has to be [0.79, 0.79+0.02] or [0.79, 0.81]. 
Numerically it is seen to converge to approximately 0.80. \hfill $\Box$

\bibliographystyle{plainnat}
\bibliography{reference}

\begin{thebibliography}{12}
\providecommand{\natexlab}[1]{#1}
\providecommand{\url}[1]{\texttt{#1}}
\expandafter\ifx\csname urlstyle\endcsname\relax
  \providecommand{\doi}[1]{doi: #1}\else
  \providecommand{\doi}{doi: \begingroup \urlstyle{rm}\Url}\fi

\bibitem[Arthur(1994)]{Arthur1994a}
W.~B. Arthur.
\newblock Inductive reasoning and bounded rationality: the {E}l {F}arol
  problem.
\newblock \emph{Am. Econ. Rev.}, 84:\penalty0 406--411, 1994.

\bibitem[Banerjee et~al.(2012)Banerjee, Mitra, and Mukherjee]{Mitra_etal;12}
P.~Banerjee, M.~Mitra, and C.~Mukherjee.
\newblock Kolkata paise restaurant problem and the cyclically fair norm.
\newblock In F.~Abergel et~al, editor, \emph{Econophysics of Systematic Risk
  and Network Dynamics}, pages 201--216. New Economic Window Series, Springer
  Verlag Italia, Milan, 2012.

\bibitem[Carter(2001)]{Carter2001}
Michael Carter.
\newblock \emph{Foundations of Mathematical Economics}.
\newblock MIT Press, 2001.

\bibitem[Chakrabarti et~al.(2009)Chakrabarti, Chakrabarti, Chatterjee, and
  Mitra]{chakrabarti2009}
A.~S. Chakrabarti, B.~K. Chakrabarti, A.~Chatterjee, and M.~Mitra.
\newblock The kolkata paise restaurant problem and resource utilization.
\newblock \emph{Physica A}, 388:\penalty0 2420 -- 2426, 2009.

\bibitem[Chakraborti et~al.(2015)Chakraborti, Challet, Chatterjee, Marsili,
  Zhang, and Chakrabarti]{Chakraborti2015}
A.~Chakraborti, D.~Challet, A.~Chatterjee, M.~Marsili, Y.-C. Zhang, and B.~K.
  Chakrabarti.
\newblock Statistical mechanics of competitive resource allocation using
  agent-based models.
\newblock \emph{Physics Reports}, 552:\penalty0 1–25, 2015.

\bibitem[Challet and Zhang(1997)]{Challet1997}
D.~Challet and Y.-C. Zhang.
\newblock Emergence of cooperation and organization in an evolutionary game.
\newblock \emph{Physica A}, 246:\penalty0 407, 1997.

\bibitem[Challet et~al.(2004)Challet, Marsili, and Zhang]{Challet2004}
D.~Challet, M.~Marsili, and Y-C. Zhang.
\newblock \emph{Minority games: interacting agents in financial markets}.
\newblock Oxford Univ. Press, Oxford, 2004.

\bibitem[DeGroot(1974)]{DeGroot_74}
M.~H. DeGroot.
\newblock Reaching a consensus.
\newblock \emph{Journal of the American Statistical Association}, 69\penalty0
  (345):\penalty0 118--121, 1974.

\bibitem[Fogel et~al.(1999)Fogel, Chellapilla, and Angeline]{Fogel1999}
D.~B. Fogel, K.~Chellapilla, and P.~J. Angeline.
\newblock Inductive reasoning and bounded rationality reconsidered.
\newblock \emph{{IEEE} transactions on evolutionary computation}, 3
  (2):\penalty0 142, 1999.

\bibitem[Ghosh et~al.(2010)Ghosh, Chatterjee, Mitra, and
  Chakrabarti]{Ghosh2010}
A.~Ghosh, A.~Chatterjee, M.~Mitra, and B.~K. Chakrabarti.
\newblock {Statistics of the Kolkata Paise Restaurant problem}.
\newblock \emph{New J. Phys.}, 12\penalty0 (7):\penalty0 075033, 2010.

\bibitem[Jackson(2010)]{Jackson_network_10}
M.~Jackson.
\newblock \emph{Social and Economic Networks}.
\newblock Princeton University Press, 2010.

\bibitem[Sornette(2004)]{Sornette2004}
D.~Sornette.
\newblock \emph{Why Stock Markets Crash: Critical Events in Complex Financial
  Systems}.
\newblock Princeton Univ. Press, 2004.

\end{thebibliography}
\end{document}